# 断層の成長

## －シミュレーションを通して再現，予測される断層の変化－

ＢＲＧＭ* 青地秀雄

産業技術研究所活断層研究センター** 安藤亮輔

# Numerical Simulation on Faulting: Microscopic Evolution, Macroscopic Interaction and Rupture Process of Earthquakes

Hideo Aochi

BRGM, Land Planning and Natural Risk Division

Ryosuke Ando

Advanced Industrial Science and Technology (AIST), Active Fault Research Center

* BRGM/ARN/RIS, 3 avenue Claude Guillemin, Orléans Cedex 2, F45060, France

** 〒305-8567 茨城県つくば市東１－１－１ / Central 7, Tsukuba, Ibaraki 305-8567, Japan

**Abstract**

We review the recent researches of numerical simulations on faulting, which are interpreted in this paper as the evolution of the state of the fault plane and the evolution of fault structure. The theme includes the fault constitutive (friction) law, the properties of the gauge particles, the initial phase of the rupture, the dynamic rupture process, the interaction of the fault segments, the fault zone dynamics, and so on. Many numerical methods have been developed: boundary integral equation methods (BIEM), finite difference methods (FDM), finite or spectral element methods (FEM, SEM) as well as distinct element methods (DEM), discrete element methods (again DEM) or lattice solid models (LSM). The fault dynamics should be solved as a complex non-linear system, which shows multiple hierarchical structures on its property and behavior. The researches have progressively advanced since the 1990's both numerically and physically thanks to high performance computing environments. The interaction at small scales is modeled to provide a large scale property of the fault. The dynamic rupture has been actively studied especially for the effect on the fault geometry evolution or due to the existed fault structure. The (quasi-)static and the initial processes of the fault movement have been also explored in a seismic cycle. The effect of fluid or heat has been also taken into account in the mechanics. All these efforts help us to understand the phenomena and the unified understanding (simulation) over different spacio-temporal scales is more and more expected.

Key words: fault constitutive law, gauge particles, fault state,

dynamic rupture, fault geometry

## §1. はじめに

"The mechanics of earthquakes and faulting"が出版された
のが 1990 年 [C. H. Scholz (1990; 第 2 版 2002)]. 日本語
訳「地震と断層の力学」もその 2 年後に出版され [ショ
ルツ (柳谷訳, 1993)], 文字通り地震と断層の力学を学ぶ
学生の間では当時標準的な教科書となった. これを見直
してみると, 凝着力, すべり (速度) 弱化摩擦法則, 破
壊核の形成, 断層形態, 断層帯のレオロジー, 地震サイ
クルなどの概念はすでに提案されていたことがわかる.
もちろん, Scholz 教授は実験を主とする研究者であるが,
数値計算 (シミュレーション) に関する項目は多くない
ことに気づく.

シミュレーションという点で, 例えばその地震学的意
義 (地震波放射にからんだ断層運動過程) からすれば,
既に Aki and Richards (1980; 第 2 版 2002) に 60-70 年代
の進展をみることができる. 現在使われているシミュレ
ーション手法や, 断層運動の概念はこの時代に原点をな
すことがわかって面白い.

しかしながら, 「断層」は狭義の地震学の枠組みを超え
て, 地球科学, 固体力学の複合領域の対象となり, その
意味でのシミュレーションは 90 年代以降計算機(コンピ
ュータ) の進化とともに劇的に進んだように見える. い
まだ研究, 議論が進行中ということもあり, 筆者の知る
限り完全な最近のレビューはないように思えるが, それ
故にこの特集号において試みる意義があろう.

「断層の成長」というテーマに対して, 「断層面の微視
的 (ミクロな) 状態の変化」あるいは「断層形状の巨視
的 (マクロな) 変化」と解釈することにする. ミクロ,
マクロの境は厳格ではなく考えているスケールにも依存
し, この間をメゾスコピックと表することもできる. い

ずれにせよ重要なのは，これからみるように断層が階層
構造をなす非常に複雑な対象であるということである．
この断層の状態変化は動的（地震時）にも静的にも起こ
るであろう．これらをキーワードにシミュレーションと
いう観点から 90 年代以降の研究を中心にレビューして
みたい．なお，断層レオロジーの一般的レビューに関し
ては本特集号第 10 章「岩石実験・レオロジー」を，断層
に関するシミュレーションでも特にプレート境界で起こ
るような再来性の強い地震については第 7 章「地震発生
の数値シミュレーション」の他のレビューを参考にして
いただきたい．「断層の成長」という物理的な理解に加え
て，手段としてのシミュレーション手法の開発も重要な
研究テーマであるので，最初に簡単なレビューを加える
ことにする．

## §2. 断層運動のシミュレーション手法

　最初にシミュレーション手法の発展を見ておこう．大
別して，連続体中の食い違い問題，あるいはミクロな物
体（質点）同士の相互作用問題として扱われうる．後者
も集合的には連続体の特徴をしめす．

### 2.1 連続体モデル

　連続体中で断層運動をシミュレーションする手法は，
（半）解析的手法，純粋な数値的手法が考えられる．両
者とも 60-70 年代に多くの先駆的仕事を見ることができ
るが，90 年代の計算機の進歩により精度と複雑な状況で
の解を求めてさらに発展した．以下，動的な問題を中心
に述べる．静的問題に関しては本号の他のレビューを参
照願いたい．なおここではあえてこの時代の博士論文を
中心に流れをたどる．これは投稿論文では数値計算手法
が十分に述べられていない，あるいは手法という地道な

テーマが論文にすらされていないという現実がある．幸い，インターネットの発展で大部分の博士論文は電子的に入手可能である．応用としては同名著者の論文（本稿の参考文献にも多数あり）を拾っていただきたい．

（半）解析的手法は連続体の解が存在し，変形が線形重ねあわせで書ける範囲で有効であり長い歴史がある．精度がよい反面，複雑なジオメトリーでは解が複雑になるのと，積分方程式のゆえに積分領域が時空間的に大きくなると計算時間がかかる．動的には無限媒質中の非平面断層における境界積分法 (BIEM, BEM) が大きく発展 [Tada (1996), Kame (1998), Aochi (2000), Aochi et al. (2000), Tada and Madariaga (2001), Tada (2006)] し，半無限媒質体への拡張 [Zhang and Chen (2006)] や高速に計算する努力などがされた [Ando et al. (2007)].

純粋な数値的手法でも，差分法 (FDM) は複数のグループによって開発が続けられた．ここでも非平面断層の動的問題を扱う手法が提案されているが [Harris and Day (1993), Kase and Kuge (1998), Kase (2000), Cruz-Atienza (2006), Cruz-Atienza et al. (2007)]，断面面上の境界条件の扱い方については常に問題が提起されていることを指摘しておきたい [Madariaga et al. (1998), Andrews (1999)].
有限要素法 (FEM) ももう一つの有力な手法であるが，動的問題に関しては差分法より計算が数段重いのと，静的問題に対しては地球内部変形を模するモデル領域端の扱い方に問題があるように思われる．いずれにしても領域の取り方に関しては前述の手法より自由が利き，不均質媒質，複雑な断層構造など現実に近い設定で断層運動がシミュレーションされる（動的な場合には，例えば Oglesby (1999)， Aagaard (2000)，Badea et al.(2008)．ただ計算結果はあまり定量的には議論されていない．動的

問題に関して，有限要素法の欠点である計算量を少なくするようなスペクトラル要素法 (SEM) が開発された [Ampuero, 2002; Festa, 2004]．解が高次関数で記述され，ローカルに解かれるおかげで断層運動の記述が高精度の割りに簡便になされるのは特筆すべきことである．また，X-FEM と呼ばれる，有限要素内の内挿に断層により生じる変位場の特性(断層面での不連続や先端の応力特異性)を考慮した基底関数を用いて，要素形状によらず任意形状の断層を扱える手法も開発されている[Sukumar, et al., 2000]．有限要素法の変分原理を使わずに，要素内でのエネルギー保存から解く有限体積法 (FVM) も提案されている [Benjemaa et al., 2007]．要素内での基底関数の次数によって解の精度を向上できる点や，要素間の不連続を原理的には苦にしない点で有限要素法と同様な長所がある．これら数値手法のすべての努力は断層面での境界条件を正確にいれることと，本質的に不連続な断層運動が引き起こす数値不安定を抑えることにあると思われる．

## 2.2　格子・粒子モデル

　断層のように十分破砕されていると思われる物を対象に，個々の物質，粒子同士の接触・相互作用の集合としてマクロな挙動を表現する手法もあり，さまざまな物理現象に適用されている．断層に対して，このような手法は，媒質の粒子性の強い断層ガウジ帯や，凹凸のある断層面，マイクロクラック生成の再現に適している．個々の相互作用を時々刻々解くという原理上，大規模計算が必然である．予め格子を空間に固定し，格子間の相互作用（連続ないし不連続）を解く格子モデル（LSM ないし DiscreteEM）と呼ばれる手法 [e.g. Sakaguchi and Mühlhaus (2000), Dalgeur et al. (2003)] や，格子を仮定せ

ずに粒子を配置し自由に運動させる粒子法ないし個別要素法（DistinctEM）という手法 [e.g. Morgan and Boettcher (1999)] がある．いずれの場合にも個々の相互作用には簡単な物理法則が与えられる場合が多く，例えば，粒子間の接合時にはフック則による弾性，結合が切れれば単純な摩擦則などが与えられる．しかしながら手法の命名には注意が必要であり，各グループがどのような設定で行っているか詳細を見る必要がある．粒子同士を格子で結べば格子法とみなせるが，接合が切れたのちに運動を許せば粒子法に近い [Mora and Place (1994, 1999)]．最近では，Hori et al. (2005)により，FEM に DiscreteEM の要素を取り入れた手法が開発され，FEM-βと呼ばれる．

　地球内部のように封圧が十分かかったような環境下では，粒子同士が起こす相互作用の自由度は低いようにも思えるが，断層ガウジのシミュレーションでは，断層が室内実験で得られた結果ほどは強くないという主張を支持することができる[Mora and Place (1999)]．また，DEM による計算でも，多数の粒子を結合させて多数の大きな粒子を構成し，それを充填した層を剪断変形させることで，ガウジ物質の破砕過程を考慮した計算も出てきた [Abe and Mair (2005)]．

### §3.　断層面の状態の変化

　断層面の状態の変化とその影響についてはさまざまなモデルが提唱された．本質的には，動的破壊時の挙動は弱化過程であり，それ以外では強化過程であるということ，それぞれの過程が何らかのスケール依存性を持つということである．

### 3.1 断層強度の変化

断層強度の変化自体を物理的にモデル化するのはいまだ難しい．例えば，広くシミュレーションに用いられる rate and state-dependent 摩擦法則本特集号の他のレビューを参考のこと）の state（状態）の実態は，Ruina (1983) で定式化された段階で定義されていない．断層強度の変化を担うこの「状態」が物理的に何であるかは重要な問いであり，熱化学過程とも結び付けられる [e.g. Nakatani (2001)]．力学的な一つのアイデアとしては，図 1 に示すように，断層トポロジーの変化と結びつけることであり，弱化過程はトポロジーの磨耗 [Matsuura et al. (1992)] に，強化過程は表面の凝着とトポロジーの回復 [Aochi and Matsuura (2002)] によるとすることができる．このモデルの長所は，弱化過程の特徴的すべり量（一般に Dc と称される）のスケール依存性をトポロジーに結び付けて説明できることである．力学的シミュレーションをもとに，ミクロな弱面の集合がマクロな Dc を導くということも言われている [Yamashita and Fukuyama (1996), Campillo et al. (2001), Ide and Aochi (2005)]．陽に仮定するかどうかは別にしてこのパラメーターDc のスケール依存性は非常に重要なテーマである（次章も参照のこと）．

別の見方として，断層に存在するガウジが主たる役割を担うと考えられることもある．一対のガウジ粒子間では単純な摩擦法則が多くの場合仮定されるが，多くの粒子の噛み合わせにより集合的に複雑な断層面の状態の変化が生み出される [Morgan and Boettcher (1999), Place and Mora, 2000]．

純粋に力学的考察だけではなく，流体や熱を考慮した断層の物理化学的モデルも盛んである[Suzuki and Yamashita (2006)]．多くの場合，流体は封圧を下げる力学

的効果 [Miller et al. (1996), Yamashita (1997)]，熱は破壊の
エネルギー消費にからんで力学と独立に強度を下げる
（化学的効果）[Shaw (1995), Andrews (2002), Rice (2006)]
と一般に考えられる．

## 3.2 断層の状態変化がもたらす地震現象

一方，これらの断層面の変化はさまざまな地震現象を
もたらす．断層強度の弱化過程とその空間不均質が，外
部応力の高まりに応じて準静的な破壊核形成過程がシミ
ュレーションされた [Matsu'ura et al., 1992; Dieterich,
1992]．この準静的過程から地震の動的過程への初期破壊
過程，加速過程についても実験や解析結果とともに，シ
ミュレーション可能になり [Shibazaki and Matsuura,
1992, 1998; Campillo and Ionescu, 1997; Lapusta et al.,
2000]，実験や地震データ解析結果との比較が可能にな
った．さらに人為的に断層面の状態をリセットすること
なく，変化が常に一つの方程式系で記述されている場合
の地震サイクルのシミュレーションが可能になり，地震
活動の複雑性と非地震性すべりなどが議論されるように
なった [Rice (1993), Ben-zion et al. (2003)]．

## 3.3 断層面の動的破壊現象

地震波を放射するという意味で地震学にとって非常に
重要な断層面上における動的破壊現象を詳しく振り返っ
てみよう．前節までで見たように，断層強度の変化がさ
まざまに議論されるに伴って，シミュレーションを通し
てその地震学的意義が考えられるようになった
[Cochard and Madariaga (1994), Ben-Zion and Rice (1997),
Fukuyama and Madariaga (1998)]．しかし結果として，rate
（すべり速度）依存性が内包されていようとも，マクロ

にはすべり弱化（slip-weakening）式の断層構成則で記述されるという結果に戻ってきた [Bizzarri et al., 2001]．地震破壊パラメーターの詳細が地震波形の解析から考察されるようになったこともあり，この結果は妥当であろう [Ide and Takeo, 1997; Mikumo et al., 2003]．

　そして大規模計算が 90 年代半ばには手軽に可能になったこともあり，破壊過程の不均一性に関するシミュレーションがますます盛んになった．実際の地震の自発的動的破壊がシミュレーションされ地震波形の観測記録と比較可能になるところまできた [Olsen et al., 1997; Aochi and Fukuyama, 2002; Aochi et al., 2003]（図 2 参照）．そして特筆すべき進化は断層形状に関する問題であろう．形状自体が自発的に成長するモデルすらシミュレーション可能 [Kame and Yamashita (1999, 2003); Ando and Yamashita (2007)]（次章参照，図3及び図4）になったが，マクロな地震破壊シナリオ，中低周波の地震波放射を考える限りにおいては断層形状をマクロに固定したモデルで扱われうる．前章で挙げた BIEM, FDM, FEM などの手法を用いて，断層セグメント間の破壊の飛び移り[Harris and Day (1993, 1999), Kase and Kuge (1998, 2001)]，逆断層と地表面の相互作用 [Oglesby and Day (2001), Zhang et al. (2006)], 屈曲・分岐断層での破壊進展 [Bouchon and Streiff (1997), Tada and Yamashita (1997), Aochi et al. (2002), Kame et al. (2003), Oglesby et al. (2003)] などが精力的に扱われた．これらによるとアスペリティー，バリアなどの概念や，破壊の開始・停止，加速・減速などは断層形状の複雑性からある程度説明でき，また破壊シナリオは断層形状の小さな非規則性によって大きく変わることがわかる．

　さまざまな実験，観測をもとにシミュレーションが可

能になったことなどから，破壊エネルギー，地震エネルギー，摩擦熱のエネルギー収支の問題も再び取り上げられている．地震学的には，観測される地震波のエネルギーが断層のどの部位からどれほど放射されたかが知りたいわけだが，ときに「エネルギー」のとらえ方が異なるようである [Ide (2002), Favreau and Archuleta (2003), Fukuyama (2005), Kanamori and Rivera (2006), Cocco et al. (2006), Madariaga et al. (2006)]．破壊力学的には断層面上のある一点（単位面積あたり）でのエネルギーが議論されるのに対して，地震学的にはマクロな積分量であることにことに注意しなければならない [Kanamori and Rivera (2006)]．

## §4. 断層形状の巨視的な変化

### 4.1. 観測事実とモデル化

前節では，断層滑り面の微視的振る舞いが，巨視的構成則としてどのように表現されるかを見た．本節では，視点を変えて，断層の巨視的形状や断層帯の内部構造の成長，進化に関する数値計算を用いたモデル化，シミュレーションの研究をレビューする．

まず，断層の成長，進化に関して概ね受け入れられた観察事実，経験則をまとめておく．長大な断層系は，一度の地震破壊で形成されるのではなく，幾度もの地震の繰り返しに伴い，成長するものと考えられている [Scholz (2002), pp. 101]．例えば，Cowie and Scholz (1992) によれば，断層の地表変位のプロファイルより，孤立した断層の先端が一回の地震に伴い伸びる長さはごく僅かである．Lockner, et al. (1991)などの室内実験では，無垢の岩石が載荷とともに即座になめらかな一枚の破壊面を作ることはなく，まず微小破壊が試料全体で生じ，それ

が徐々に局在化し，最終的にその面が動的に破壊すると
いう経過をたどる．地質学的観察も，近接した小断層が
結合して，断層が長大化することを示唆している[Segall
and Pollard (1983)].

　また，断層は変位を累積し成熟すると単純化していく
ように見える [Ben-Zion and Sammis (2003)]. 例えば，断
層は成熟するにつれて，個々の断層セグメントは長くな
り平坦化するようである [Stirling et al. (1996),
Wesnousky (1988)]. 断層帯の内部構造を見ると，若い断
層は多くの滑り面から構成されるのに対し，成熟すると
主たる滑り面が顕わになり，ガウジ層が発達するととも
に，その周辺の破砕帯も成長する [Chester et al. (2005),
Di Toro and Pennacchioni (2005), Scholz (1987), Vermilye
and Scholz (1998)].

　断層の成長の問題は，地質学的時間スケールと地震学
的時間スケール，断層スケールと微小亀裂のスケールな
どと，大小の時空間スケールの交錯するやっかいな問題
であり，現在の観測手法はその過程を明らかにするのに
十分ではない．数値計算によるモデル化は，地質調査や
室内実験が苦手とするそのような対象を，統一的に理解
するのに有効である．

## 4.2.断層形状の進化

　まず，断層の結合過程と形状の進化を直接的にモデル
化した研究を概観する. Renshaw and Pollard (1994) や Du
and Aydin (1995) は，慣性項を無視した静的 BEM を用い
て，断層は先端での応力が決める最適面の方向に進展す
ると仮定して，断層の成長をシミュレーションした．系
を準静的に歪ませることで，近接する断層は応力を介し
た相互作用により，引き付けあい互いに結合することを

示し，それは，野外観察でよく見られる多数の断層セグ
メントが Jog により連鎖している様子をよく再現してい
る．Ando et al. (2004) は，断層の結合は，地震時の動的
破壊に伴い生じると考え，動的 BEM を用いて，その効
果を考察した．その結果，破壊伝播速度が速いほど，断
層間の結合が促進されることを示した．この現象は，
Kame and Yamashita (1999) が明らかにした，孤立した断
層の進展に与える，動的破壊伝播の効果と同様のもので
ある（図 3 参照）．また，地震サイクルの効果を考慮する
ため，動的-準静的統合型の BEM [Ando et al. (2007)]を用
いた試みも始まっている．

　破壊面形状を表現するには，個別要素法（DEM）を使
った計算もあるが，強い不均質物質の破壊やガウジ層内
部の変形など最小単位が粒状体である性質を生かしたも
のが多い．無垢の岩石の破壊実験の様子は，Scott (1996)
などで再現されている．Morgan and Boettcher (1999) は，
DEM でリーデル剪断面の形成と滑りに伴う剪断歪みの
局在化を再現している．

　Lyakhovsky et al. (1997) は，断層を変位の食い違いと
境界条件という取り扱いではなく，ダメージレオロジー
の理論を用いて，断層帯をバルクでの弱領域として表現
し，準静的枠組みで，断層の進化をモデル化している．
歪み弱化の過程が，塑性歪みの局在化，すなわち断層を
形成することを示している．Lyakhovsky et al. (2001) で
は，弾性，粘弾性の二層構造をモデルに取り入れた．
Shibazaki et al. (2007) は，上部地殻の弾性と下部地殻の
弾塑性変形を考慮したモデルを有限要素法（FEM）で計
算し，下部地殻に局在化した塑性変形領域が生じ，上部
地殻の断層に応力集中させること示した．

### 4.3.断層サイズ分布の進化

次に，地質学的な時間スケールの断層成長にともなう，統計的性質の変化を扱った研究を紹介する．統計的性質の解明は，断層や地震破壊領域の長さ分布に関して進んでいる．これらのモデルは，系の振る舞いを長時間計算する必要性から，弾性体の振る舞いや摩擦則，断層形状をそれぞれの方法で単純化して表現している．

地震をいくつも繰り返すような長期間の振る舞いの考察は，Bak and Tang (1989) などのセルラオートマトン，Carlson and Langer (1989) などの一次元ブロックバネモデルで始まり，グーテンベルグリヒター則を再現したことで有名である．ただしこれらのモデルは，非常に抽象的であり実際の地震現象との対応は明瞭でない．これらのモデルが，断層が同一面上に並ぶ（coplanar）正の相互作用，最近接要素間の相互作用に限られていたのを，Cowie et al. (1993) は，断層が同一面にない場合（non-coplanar）に生じる負の相互作用や連続弾性体のもつ遠隔相互作用も考慮したモデル化を，準静的枠組みで行った．彼女らの計算では，始め乱雑に微小破壊が系の至る所で生じていた状態から，徐々に活動が系の幅の大部分を占めるような大断層に局在化する状態に遷移することが示されている．この局在化は，まさに大きな断層の及ぼす負の相互作用により，小さな断層の活動が抑制された結果である．

Spyropoulos et al. (2002) は，準静的な差分法（FDM）を用いて，Mode III の変形条件下で．サイズ分布とその進化をより詳細に考察した．断層の摩擦強度について，Cowie et al. (1993) が，地震時の急激な弱化とその後の地震前レベルへの急激な回復を仮定していたのに対し，この研究では滑り弱化を考慮して，滑りに合わせて断層が

弱化して成熟する過程を導入した．系を徐々に歪みを与えて駆動した結果，加えた歪みが十分に小さい時には，初期に与えた強度の不均質を反映した個々の小断層が指数関数分布で卓越し，中間程度の歪みに達した時はベキ乗則に従う断層分布が生じ，さらに歪みが加わると系のサイズ近くに達し成長が飽和した断層が支配的な指数関数分布になるという，異なる状態に遷移することを示した．彼らはこの結果を，野外での観察結果に適用し，累積的な変位の違いにより，観察される指数関数分布とベキ乗則分布が説明されると主張した．Shaw (2004) は，Spyropoulos et al. (2002) のモデルを，慣性項を考慮した動的モデルに拡張し，波動生成と動的な破壊伝播を考慮した．これにより，複数の断層セグメントを乗り越える地震イベントが再現されている．Shaw (2006) は，地震サイクルの中で形成された断層形状の効果で，地震の破壊開始点が固定化される傾向のあることを統計的に示し，地震ハザード予測に有用だと主張している．

### 4.4.動的断層成長と断層面外ダメージ，分岐断層

　最近になり，地震のエネルギー収支や断層帯の内部構造の効果が注目される中で，地震時にこの内部構造がどのように進化するかを考察した研究が行われるようになっている．

　Yamashita (2000) は，大量の微少亀裂の生成を弾性係数の低下でモデル化し，動的破壊伝播に伴う破壊先端での応力集中のために，この応力集中領域もしくは破損領域（Breakdown zone）などと呼ばれる領域で，微少亀裂が生成される過程をシミュレーションした．これは野外観察で主たる滑り面周辺に存在する破砕帯の生成モデルである．このような現象は，断層面外ダメージ（off-plane

damage）とも呼ばれる．Andrews (2005) は，断層周辺の媒質に降伏弾塑特性を仮定して，同様の過程により断層滑り面周辺の塑性変形領域が形成されることを示した．彼は，さらに主断層面上で破壊が進展することで消費されるエネルギーと，面外の塑性変形で散逸されるエネルギーの総和が，巨視的な破壊エネルギー（Breakdown work）であると主張し，それが破壊の伝播距離に比例することを示した．このスケーリング則は，破砕帯の幅が破壊伝播距離に比例するためで，巨視的構成則パラメーターが，下位スケール階層の幾何学的構造と密接に結びついている好例である．このようにスケールする破壊エネルギーはもはや断層に固有の物性値ではない．

　Ando (2005)，Ando and Yamashita (2007) は，上記のモデルがバルクの特性として破砕帯を表現したのに対して，主断層の周辺に変位の食い違いによる多数の分岐断層を生じさせるモデル化を行った（図 4）．彼らは，分岐断層の長さは，主断層上の破壊伝播距離がある臨界値を超えるかどうかで大きく異なり，それ以下では，Andrews (2005) と同様のスケーリングに従い，主断層より圧倒的に短い（メゾ分岐）が，それ以上では，分岐断層が自発的に成長し主断層と同程度にまで成長する(マクロ分岐)し，単純なスケーリングが破れる可能性のあることを示した．彼らは，従来曖昧なスケールの定義に関し，局在化した滑り面の厚さ以下のミクロと断層長さのマクロの中間に，断層帯の厚さをメゾとして導入し力学的に考慮することを提案している．Ando (2005) でも，Andrews (2005) と同様の破壊エネルギーに関するスケーリングも導いている．

　Dalguer et al.(2003) は，DEM で三次元媒質内での，断層周辺の微小開口亀裂の生成をシミュレートし，上記 2

次元の場合と同様の断層帯幅のスケーリングを示した．Ben-Zion and Shi (2005) は，2次元弾性媒質で弾性波速度境界に断層を埋め込み，Andrews (2005) と同様の弾塑性変形を考慮し，動的破壊を計算した．その結果，速度境界の効果で破壊の伝播方向に指向性が生じるため，断層の片側に破砕帯が成長しやすいと主張している．

### 4.5. 断層成長における破壊基準の問題

最後に，破壊基準という古典的問題を再考しておきたい．Mode II の剪断破壊は，剪断応力依存性のみでモデル化されることもある[例えば，Du and Aydin (1995), Kame and Yamashita (1999), Ando et al (2004)]が，通常のCoulomb の破壊基準を考慮すれば，断層端での剪断応力の対称性に対し，法線応力は反対称であるために，その走行方向には（準静的にも）成長できず [Scholz (2002), pp. 31] 一定方向に屈曲しながら進展せざるを得ない．一方で，無垢の岩石の破壊実験で生じる最終破断面のような剪断破壊面の向きは，巨視的には直線的で，内部摩擦角の方向としてCoulomb基準でよく説明されるとされる．この矛盾は，既存のモデルで力学的に十分に解決されておらず，また数値的にも解かれたとは言えないように思われる．[Scholz (2002), pp. 110] が指摘するように，Mode II の破壊面は，先だって形成された Mode I の破壊面をなぞると考えるのは一般性に欠ける．破壊面が階層的構造であるならば，ある特定のスケールでの面だけがCoulomb 基準に従うと考えるのも，一般的には不自然であろう．前述の DEM による計算でも，粒子性（粒子の配置と粒径の効果）に影響され，系全体より十分小さな亀裂の先端の応力場とその進展方向は明瞭ではないし，何より連続弾性体との関係も厳密には明らかではない

[Hori et al. (2005)].形成される剪断破壊面が膨大な微小破壊面の結合の結果であることを考えれば，従来の手法の延長線上で，このような過程を再現するには，より大規模で困難な数値計算を行う必要がある．したがって，新たなモデル化手法の開発も含めて，Mode II の破壊基準を確立することは，断層成長の問題に関する基本的だが今もって挑戦的課題であろう．

### §5. 今後の展望とまとめ

1990 年以降の研究を中心に，「断層の成長」に関するシミュレーションという観点でレビューを行ってきた．筆者らは「断層面の状態の変化」と「断層形状の巨視的な変化」という捉え方をした．いくつかキーワードが存在した．その一つは，複雑系の科学であろう．断層は力学的に複雑であるだけでなく，物質科学的，地質学的にも複雑であり，因果関係を定量的に明らかにするためには大規模シミュレーションが必要とされる．もう一つは，時空間的なスケールの問題であろう．断層は地震のように非常に短い時間でその状態を劇的に変えるが，地質学的時間では異なったレオロジーで成長する．空間的には，断層破砕帯のような狭い領域で起こる現象が，マクロに断層を見たときにどのように記述されるか，さらにそのマクロな断層群はどのように相互作用をしながら成長するか，といったように問題が階層的に連なっており，物理法則は必ずしも明らかではない．このように異なる時空間で起こる断層現象を無理なく統一的に理解するのが目的であり，観測や実験と統合化したシミュレーションが求められる．

## 文献

Aagaard, B., 2000, Finite-element simulations of earthquakes, PhD Thesis, California Institute of Technology, USA.

Abe, S., and K. Mair, 2005, Grain fracture in 3D numerical simulations of granular shear, Geoph. Res. Lett., **32**, L05305, doi: 1029/2004GL022123

Aki, K. and P. Richards, 1980, Quantitative Seismology; theory and methods, W. H. Freeman and Co.

Ampuero, J.-P., 2002, Etude physique et numérique de la nucléation des séismes, PhD Thesis, University of Paris VII, France.

Ando, R., 2005, Development of Efficient Spatio-temporal Boundary Integral Equation Method and Theoretical Study on Dynamics of Fault Zone Formation and Earthquake Ruptures, University of Tokyo, Japan.

Ando, R., N. Kame, and T. Yamashita, 2007, An efficient boundary integral equation method applicable to the analysis of non-planar fault dynamics. Earth Planets Space, **59**, 363-373.

Ando, R., T. Tada, and T. Yamashita, 2004, Dynamic evolution of a fault system through interactions between fault segments. J. Geophys. Res., **109**, doi:10.1029/2003JB002665.

Ando, R., and T. Yamashita, 2007, Effects of mesoscopic-scale fault structure on dynamic earthquake ruptures: Dynamic formation of geometrical complexity of earthquake faults, J. Geophys. Res., **112**, doi:10.1029/2006JB004612.

Andrews, D. J., 1999, Test of two methods for faulting in finite-difference calculations, Bull. Seismol. Soc. Am., **89**, 931-937.

Andrews, D. J., 2002, A fault constitutive relation accounting for thermal pressurization of pore fluid, J. Geophys. Res., **107**, doi:10.1029/2002JB001942.

Andrews, D. J., 2005, Rupture dynamics with energy loss outside the slip zone. J. Geophys. Res., **110**, B101307, doi:10.1029/2004JB003191.

Aochi, H., 2000, Theoretical studies on dynamic rupture propagation along a 3D non-planar fault system, PhD thesis, University of Tokyo, Japan.

Aochi, H., E. Fukuyama and M. Matsu'ura, 2000, Spontaneous Rupture Propagation on a Non-planar Fault in 3D Elastic Medium, Pure appl. Geophys., 157, 2003-2027.

Aochi, H. and E. Fukuyama, 2002, Three-dimensional nonplanar simulation of the 1992 Landers earthquake, J. Geophys. Res., **107**, doi.10.1029/2000JB000061.

Aochi, H., E. Fukuyama and R. Madariaga, 2003, Constraints of Fault Constitutive Parameters Inferred from Non-planar Fault Modeling, Geichemistry, Geophysics, Geosystems, **4**, 10.1029/2001GC000207.

Aochi, H., R. Madariaga and E. Fukuyama, 2002, Effect of normal sterss during rupture propagation along nonplanar faults, J. Geophys. Res., **107**, 10.1029/2001JB000500.

Aochi, H. and M. Matsu'ura, 2002, Slip- and time-dependent fault constitutive law and its significance in earthquake generation cycles, Pure appl. Geophys., **159**, 2029-2044.

Bak, P., and C. Tang, 1989, Earthquakes as a Self-Organized

Critical Phenomenon. J. Geophys. Res., **94**, 15635-15637.

Badea, L., I. R. Ionescu and S. Wolf, 2008, Schwarz method for earthquake source dynamics, J. Comp. Phys., **227**, 3824-3848.

Benjemaa M., N. Glinsky-Olivier, V. M. Cruz-Atienza, J. Virieux, and S. Piperno, 2007, Dybnamic non-planar crack rupture by a finite volume method, Geophys. J. Int., **171**, 271-285.

Ben-Zion, Y., M. Eneva and Y. Liu, 2003, Large earthquake cycles and intermittent criticality on heterogeneous faults due to evolving stress and seismicity, J. Geophys, Res., **108**, doi:10.1029/2002JB002121.

Ben-Zion, Y., and C. G. Sammis, 2003. Characterization of fault zones. Pure Appl. Geophys., **160**, 677-715.

Ben-Zion, Y., and Z. Q. Shi, 2005, Dynamic rupture on a material interface with spontaneous generation of plastic strain in the bulk. Earth Planet Sc Lett, **236**, 486-496.

Ben-Zion, Y. and J. R. Rice, 1997, Dynamic simulations of slip on a smooth fault in an elastic solid, J. Geophys. Res., **102**, 17771-17784.

Bizzarri, A., M. Cocco, D. J. Andrews and E. Boschi, 2001, Solving the dynamic rupture problem with different numerical approaches and constitutive laws, Geophys. J. Int., **144**, 656-678.

Bouchon, M. and D. Streiff, 1997, Propagation of a shear crack on a nonplanar fault: A method of calculation, Bull. Seism. Soc. Am., **87**, 61-66.

Campillo, M., P. Favreau, I. R. Ionescu and C. Voisin, 2001, On the effective friction law of a heterogeneous fault, J.

Geophys. Res., **106**, 16307-16322.

Campillo, M. and I. R. Ionescu, 1997, Initiation of antiplane shear instability under slip dependent friction, J. Geophys. Res., **102**, 20363-20371.

Carlson, J. M., and J. S. Langer, 1989, Mechanical Model of an Earthquake Fault. Phys Rev A, **40**, 6470-6484.

Chester, J. S., F. M. Chester, and A. K. Kronenberg, 2005, Fracture surface energy of the Punchbowl fault, San Andreas system. Nature, **437**, 133-136.

Cochard, A., and Madariaga, R., 1994, Dynamic Faulting under Rate-Dependent Friction: Pure and Applied Geophysics, **142**, 419-445.

Cocco, M., P. Spudich and E. Tinti, 2006, On the mechanical work absorbed on faults during earthquake ruptures, in "Earthquake: Radiated energy and the physics of faulting", eds. R. Abercrombie, A. MaGarr, H. Kanamori and G. Di Toro, AGU Geophysical Monograph Series 170, 237-254.

Cowie, P. A., and C. H. Scholz, 1992, Growth of Faults by Accumulation of Seismic Slip. J. Geophys. Res., **97**, 11085-11095.

Cowie, P. A., C. Vanneste, and D. Sornette, 1993, Statistical Physics Model for the Spatiotemporal Evolution of Faults. J. Geophys. Res., **98**, 21809-21821.

Cruz-Atienza, V. M., 2006, Rupture dynamique des faille non-planaires en différences finies, PhD Thesis, University of Nice Sophia, France.

Cruz-Atienza, V. M., J. Virieux and H. Aochi, 2007, 3D finite-difference dynamic-rupture modeling along nonplanar faults, Geophysics, **72**, SM123-SM137.

Dalguer, L. A., K. Irikura, and J. D. Riera, 2003, Simulation of tensile crack generation by three-dimensional dynamic shear rupture propagation during an earthquake. J. Geophys. Res., **108**, doi:10.1029/2001JB001738.

Dieterich, J. H., 1992, Earthquake nucleartion on faults with rate- and state-dependent strength, Tectonophys., **211**, 115-134.

Di Toro, G., and G. Pennacchioni, 2005, Fault plane processes and mesoscopic structure of a strong-type seismogenic fault in tonalites (Adamello batholith, Southern Alps). Tectonophysics, **402**, 55-80.

Du, Y. J., and A. Aydin, 1995, Shear Fracture Patterns and Connectivity at Geometric Complexities Along Strike-Slip Faults. J. Geophys. Res., **100**, 18093-18102.

Favreau, P. and R. J. Archuleta, 2003, Direct seismic energy modeling and application to the 1979 Imperial Valley earthquake, Geophys. Res. Lett., **30**, 1198, doi:10.1029/2002GL015968.

Festa, G., 2004, Fault dynamics with spectral elements and slip imaging by isochrone back-projection, PhD thesis, University of Bologna, Italy.

Fukuyama, E., 2005, Radiation energy measured at earthquake source, Geophys. Res. Lett., **32**, L13308, doi:10.1029/2005GL022698.

Fukuyama, E. and R. Madariaga, 1998, Rupture dynamics of a planar fault in a 3D elastic medium: rate-and slip-weakening friction, Bull. Seism. Soc. Am., **88**, 1-17.

Harris, R. A. and S. M. Day, 1993, Dynamics of fault interaction; parallel strike-slip faults, J; Geophys. Res., **98**, 4461-4472.

Harris, R. A., and Day, S. M., 1999, Dynamic 3D simulations of earthquakes on en echelon faults: Geophysical Research Letters, **26**, 2089-2092.

Hori, M., K. Oguni, and H. Sakaguchi, 2005. Proposal of FEM implemented with particle discretization for analysis of failure phenomena. J. Mech. Phys. Solids,**53**, 681–703

Ide, S., 2002, Estimation of radiated energy of finite-source earthquake models, Bull. Seosm. Soc. Am., **92**, 2994-3005.

Ide, S. and H. Aochi, 2005, Earthquakes as multiscale dynamic ruptures with heterogeneous fracture surface energy, J. Geophys. Res., 110, B11303, doi.10.1029/2004JB003591.

Ide, S. and M. Takeo, 1997, Determination of constitutive relations of fault slip based on seismic wave analysis, J. Geophys. Res., **102**, 27379-27391.

Kame, N., 1998, Theoretical study on arresting mechanism of dynamic earthquake faulting, PhD Thesis, The University of Tokyo, Japan.

Kame, N., J. R. Rice and R. Dmowska, 2003, Effect of pre-stress state and rupture velocity on dynamic fault branching, J. Geophys. Res., **108**, doi: 1029/2002JB002189.

Kame, N. and T. Yamashita, 1999, Simulation of the spontaneous growth of a dynamic crack without constraints on the crack tip path, Geophys. J. Int., **139**, 345-358.

Kame, N. and T. Yamashita, 2003, Dynamic branching, arresting of rupture and the seismic wave radiation in self-chosen crack path modeling, Geophys. J. Int., **155**, 1042-1050.

Kanamori, H. and L. Rivera, 2006, Energy partitioning during an earthquake, in "Earthquake: Radiated energy and the physics of faulting" eds. R. Abercrombie, A. McGarr, H. Kanamori and G. Di Toro, AGU Geophysical Monograph Series, **170**, 3-13.

Kase, Y., 2000, Significance of fault geometry in earthquake rupture process, PhD Thesis, Kyoto University, Japan.

Kase, Y. and K. Kuge, 1998, Numerical simulation of spontaneous rupture processes on two non-coplanar faults; the effect of geometry on fault interaction, Geophys. J. Int., **135**, 911-922.

Kase, Y., and Kuge, K., 2001, Rupture propagation beyond fault discontinuities: significance of fault strike and location: Geophysical Journal International, **147**, 330-342.

Lapusta, N., J. R. Rice, Y. Ben-Zion and G. Zheng, 2000, Elastodynamic analysis for slow tectonic loading with spontaneous rupture episodes on faults with rate- and state-dependent friction, J. Geophys. Res., **105**, 23765-23789.

Lockner, D. A., J. D. Byerlee, V. Kuksenko, A. Ponomarev, and A. Sidorin, 1991, Quasi-Static Fault Growth and Shear Fracture Energy in Granite. Nature, **350**, 39-42.

Lyakhovsky, V., Y. Ben-Zion, and A. Agnon, 1997, Distributed damage, faulting, and friction. J. Geophys. Res., **102**, 27635-27649.

Lyakhovsky, V., Y. Ben-Zion, and A. Agnon, 2001, Earthquake cycle, fault zones, and seismicity patterns in a theologically layered lithosphere. J. Geophys. Res., **106**, 4103-4120.

Madariaga, R., J. P. Ampuero and M. Adda-Bedia, 2006, Seismic radiation from simple models of earthquakes, in "Earthauqke: Radiated energy and the physics of faulting", eds. R. Abercrombie, A. McGarr, H. Kanamori and G. Di Toro, AGU Geophysical Monograph Series 170, 223-236.

Madariaga, R., K. B. Olsen, R. J. Archuleta, 1998, Modeling dynamic rupture in a 3D earthquake fault model, Bull. Seismol. Soc. Am., **88**, 1182-1197.

Matsu'ura, M., H. Kataoka and B. Shibazaki, 1992, Slip-dependent friction law and nucleation processes in earthquake rupture, Tectonophys. **211**, 135-148.

Mikumo, T., K. B. Olsen, E. Fukuyama and Y. Yagi, 2003, Stress-breakdown time and slip-weakening distance inferred from slip-velocity function on earthquake faults, Bull. Seism. Soc. Am., **93**, 264-282.

Miller, S. A., A. Nur and D. L. Olgaard, 1996, Earthquakes as a coupled shear stress – high pore pressure dynamical system, Geophys. Res. Lett., **23**, 197-200.

Mora, P. and D. Place, 1994, Simulation of the frictional stick-slip instability, Pure Appl. Geophys. **143**, 61-87.

Mora, P. and D. Place, 1999, The weakness of earthquake faults, Geoph. Res. Lett.,. **26**, 123-126.

Morgan, J. K., and M. S. Boettcher, 1999, Numerical simulations of granular shear zones using the distinct element method - 1. Shear zone kinematics and the micromechanics of localization. J. Geophys. Res., **104**, 2703-2719

Nakatani, M., 2001, Conceptual and physical clarification of rate and state friction: Frictional sliding as a thermally activated rheology, J. Geophys. Res., 106, 13347-13380.

Oglesby, D. D., 1999, Earthquake Dynamics on dip-slip faults, PhD Thesis, Unviersity of California Santa Barbara, USA.

Oglesby, D. D., S. M. Day, 2001, Fault geometry and the dynamics of the 1999 Chi-chi (Taiwan) earthquake, Bull. Seism. Soc. Am., **92**, 3006-3021.

Oglesby, D. D., S. M. Day, Y.-G. Li and J. E. Vidale, 2003, The 1999 Hector Mine earthquake: the dynamics of a branched fault system, Bull. Seism. Soc. Am., **93**, 2459-2476.

Olsen, K. B., Madariaga, R. and R. J. Archuleta, 1997, Three-dimensional dynamic simulation of the 1992 Landers earthquake, Science, **278**, 834-838.

Place, D. and P. Mora, 2000, Numerical simulation of localization phenomena in a fault zone, Pure Appl. Geophs. **157**, 1821-1845.

Renshaw, C. E., and D. D. Pollard, 1994, Are Large Differential Stresses Required for Straight Fracture Propagation Paths. J. Struct. Geol., **16**, 817-822.

Rice, J. R., 1993, Spatio-temporal complexity of slip on a fault, J. Geophys. Res., **98**, 9885-9907.

Rice, J. R., 2006, Heating and weakening of faults during earthquake slip, J. Geophys. Res., doi:10.1029:2005JB004006.

Ruina, A., 1983, Slip instqbility qnd stqte vqriqble friction laws, J. Geophys. Res., 88, 10359-10370.

Sakaguchi, H. and H.B. Mühlhaus, 2000, Hybrid modeling of coupled pore fluid-solid deformation problems, Pure appl. Geophys., **157**, 1889-1904.

Scholz, C. H., 1987, Wear and Gouge Formation in Brittle Faulting. Geology, **15**, 493-495.

Scholz, C. H., 1990 (2002 second eds), The Mechanics of Earthquakes and Faulting, Cambridge University.

ショルツ, C.H.（柳谷俊訳）, 1993, 地震と断層の力学, 古今書院.

Scott, D. R., 1996, Seismicity and stress rotation in a granular model of the brittle crust. Nature, **381**, 592-595.

Segall, P., and D. D. Pollard, 1983, Nucleation and Growth of Strike Slip Faults in Granite. Journal of Geophysical Research, **88**, 555-568.

Shaw, B. E., 1995, Frictional weakening and slip complexity in earthquake faults, J. Geophys. Res., **100**, 18239-18251.

Shaw, B. E., 2004, Self-organizing fault systems and self-organizing elastodynamic events on them: Geometry and the distribution of sizes of events. Geophys. Res. Lett., **31**(17), doi:10.1029/2004GL019943.

Shaw, B. E., 2006, Initiation propagation and termination of elastodynamic ruptures associated with segmentation of faults and shaking hazard. J. Geophys. Res., **111**(B8), doi:10.1029/2005JB004093.

Shibazaki, B., K. Garatani, and H. Okuda, 2007, Finite element analysis of crustal deformation in the Ou Backbone Range, northeastern Japan, with non-linear visco-elasticity and plasticity: effects of non-uniform thermal structure. Earth Planets Space, **59**, 499-512.

Shibazaki, B. and M. Matsu'ura, 1992, Spontaneous

processes for nucleation, dynamic propagation, and stop of earthquake rupture, Geophys. Res., Lett., **19**, 1189-1192.

Shibazaki, B., and Matsu'ura, M., 1998, Transition process from nucleation to high-speed rupture propagation: scaling from stick-slip experiments to natural earthquakes: Geophysical Journal International, **132**, 14-30.

Spyropoulos, C., C. H. Scholz, and B. E. Shaw, 2002, Transition regimes for growing crack populations. Phys Rev E, **65**, doi: 10.1103/PhysRevE.65.056105

Stirling, M. W., S. G. Wesnousky, and K. Shimazaki, 1996, Fault trace complexity, cumulative slip, and the shape of the magnitude-frequency distribution for strike-slip faults: A global survey. Geophys. J. Int., **124**, 833-868.

Sukumar, N., Moes, N., Moran, B., and Belytschko, T., 2000, Extended finite element method for three-dimensional crack modelling, Int. J. Numer. Methods Eng., **48**, 1549-1570.

Suzuki, T. and T. Yamashita, 2006, Nonlinear thermoporoelastic effects on dynamic earthquake rupture, J. Geophys. Res., **111**, doi: 10.1029/2005JB003810.

Tada, T., 1996, Boundary Integral Equations for the Time-Domain and Time-independent Analyses of 2D Non-planar Cracks, PhD Thesis, The University of Tokyo.

Tada, T., 2006, Stress Green's functions for a constant slip rate on a triangular fault, Geophys. J. Int., 164, 653-669.

Tada, T. and R. Madariaga, 2001, Dynamic modeling of the flat 2-D crack by a semi-analytic BIEM scheme, Int. J.

Num. Methods Engin., 50, 227-251.

Tada, T. and T. Yamashita, 1997, Non-hypersingular boundary integral equations for two-dimensional non-planar crack analysis, Geophys. J. Int., **130**, 269-282.

Vermilye, J. M., and C. H. Scholz, 1998, The process zone: A microstructural view of fault growth. J. Geophys. Res., **103**, 12223-12237.

Wesnousky, S. G., 1988, Seismological and Structural Evolution of Strike-Slip Faults. Nature, **335**, 340-342.

Yamashita, T., 1997, Mechanical effect of fluid migration on the complexity of seismicity, J. Geophys. Res., **102**, 17797-17806.

Yamashita, T., 2000, Generation of microcracks by dynamic shear rupture and its effects on rupture growth and elastic wave radiation. Geophys. J. Int., **143**, 395-406.

Yamashita, T. and E. Fukuyama, 1996, Apparent critical slip displacement caused by the existence of a fault zone, Geophys. J. Int., **125**, 459-472.

Zhang, H. and X. Chen, 2006, Dynamic rupture on a planar fault in three-dimensional half space – I. Theory, Geophys. J. Int., **164**, 633-652.

Zhang, W., T. Iwata and K. Irikura, 2006, Dynamic simulation of a dipping fault using a three-dimensional finite difference method using nonuniform grid spacing, J. Goeophys. Res., **111**, doi. 10.1029/2005JB003725.

Figure Captions

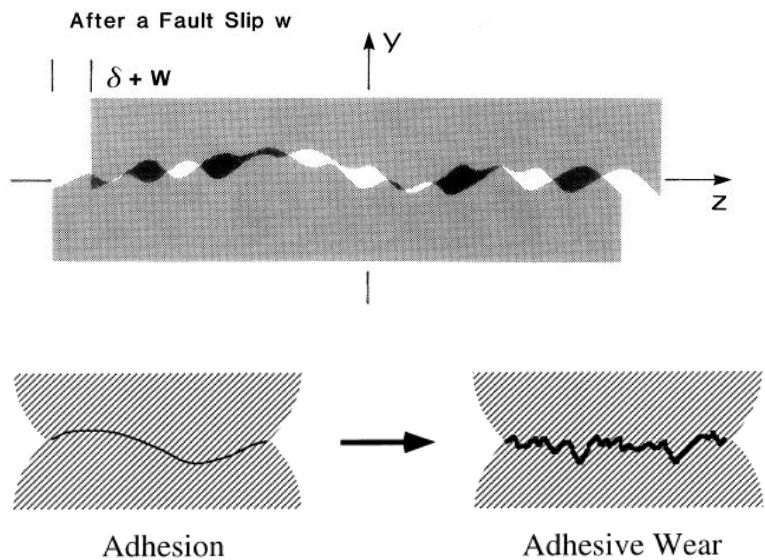

Adhesion        Adhesive Wear

Fig. 1. Schematic illustration of the change of the fault plane
during slipping and static contact. The first mechanism
is abrasion leading a weakening process with slip and
the second is adhesion for providing a recovery with
time. Figure modified after Aochi and Matsu'ura (2002).

## Dynamic simulation of the 1992 Landers earthquake

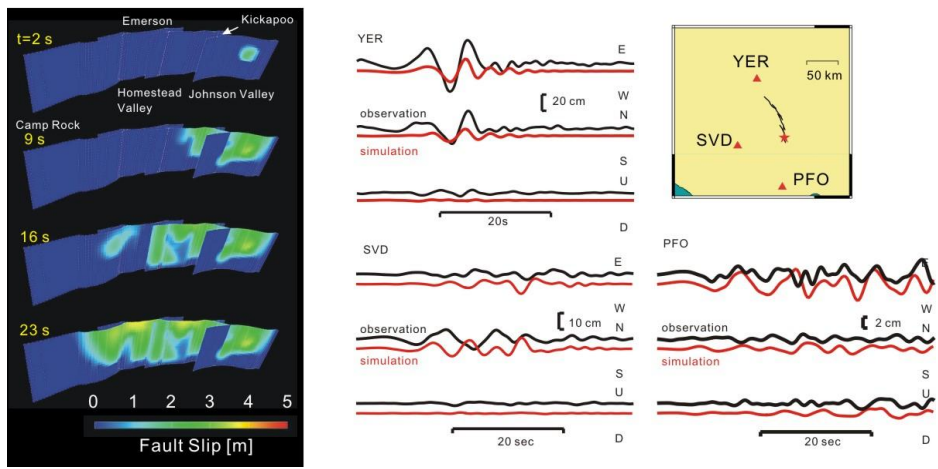

Fig. 2. Dynamic simulation of the M7.2 Landers, California, earthquake. After the simulation of the spontaneous rupture propagation along a non-planar fault system (left), synthetic seismograms are calculated for a stratified 1D structure and filtered between 0.07-0.5 Hz (right). Figure modified after Aochi et al. (2003).

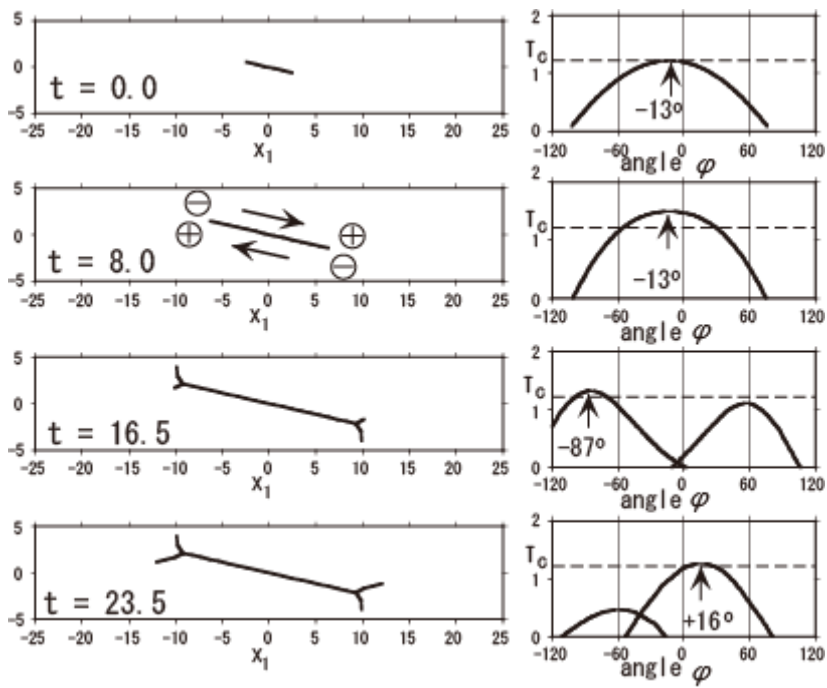

Fig. 3. Snapshots of clack growth. High speed rupture propagation causes the deviation of the hoop shear traction from its initially static distribution (right), leading the branching of the crack tips (left). Figure modified after Kame and Yamashita (2003).

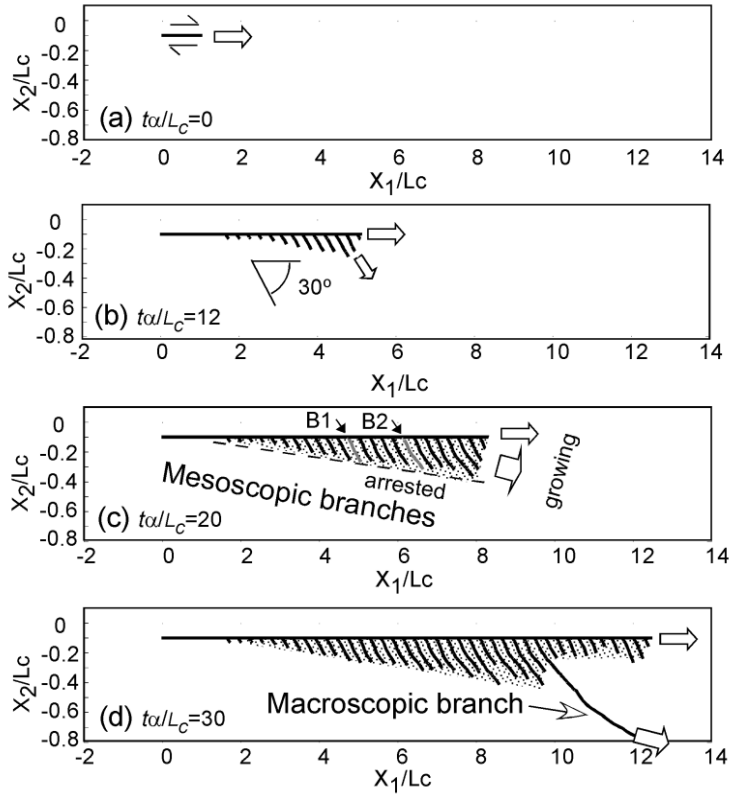

Fig.4 Snapshots of fault growth with dynamic multiple branching. The meso-branches distribute following linear scaling until (c) but the spontaneously growing macro-branch emerges as the fault attains a certain critical length in (d). The open arrows and a pair of the thin arrows indicate the direction of rupture and the slip sense, respectively. A fault zone is shaded. Figure modified after Ando and Yamashita (2007).